\documentstyle[10pt,aaspp4] {article}

\begin{document}

\title{Panchromatic Study of Nearby UV-Bright Starburst Galaxies: Implications
for Massive Star Formation and High Redshift Galaxies}

\author{Christopher J. Conselice$^1$, John S. Gallagher$^1$, Daniela Calzetti$^2$,
Nicole Homeier$^1$, Anne Kinney$^2$ }

\affil{chris@astro.wisc.edu; jsg@astro.wisc.edu; calzetti@stsci.edu; homeier@astro.wisc.edu; kinney@stsci.edu}

\altaffiltext{1}{University of Wisconsin-Madison, Department of 
Astronomy, 475 N. Charter St. Madison WI. 53706}

\altaffiltext{2}{Space Telescope Science Institute, 3700 San Martin
Dr., Baltimore, MD.}
\begin{center}

\it{Accepted for Publication in the Astronomical Journal}

\end{center}

\begin{abstract}
 
    We present a panchromatic study of nearby starburst galaxies from
the ultraviolet to the visible, including narrow band
H$\alpha$ using WIYN and HST data, to determine how star formation processes 
affect the morphology and integrated fluxes of nearby starbursts. 
We find the UV/H$\alpha$ morphology of
starbursts tend to differ, although not in a standard or predictable
manner.  From our sample of six nearby starbursts, three systems
show a good correlation between UV and H$\alpha$ fluxes, but we find
differences in UV and H$\alpha$ morphology between the other
three.  Occasionally we find systems with well defined H II regions without 
the corresponding brightness in the UV, and vice-versa.   
We discuss the likely mechanisms behind these differences which include: 
starburst ages, dust absorption, 
stellar energy ejecta through SNe and winds, as well as leakage of UV photons 
from stellar clusters.  We conclude that the large scale morphological 
features in starbursts are primarily due to both age and absorption from
a `picket fence' dust distribution.  
  
    We further demonstrate the similarity and differences between 
these nearby starbursts and high redshift star forming
galaxies.   The overall morphology of our sample of starbursts
changes {\em little} between UV and visible wavelengths.  If high redshift
galaxies are similar to these starbursts, their morphologies should 
change little between rest-frame UV and optical.  We also show that FIR and 
UV spectral energy distributions and slopes can be used to determine large 
scale morphological features for extreme starbursts, with the steepest
FIR slopes correlating with the most disturbed galaxies.

\end{abstract}


\section{Introduction}

    Starbursts are typically defined as galaxies where star formation
rates are substantially enhanced, with a significant portion of the light 
output originating from distinct star 
forming regions (Weedman et al. 1981); or as galaxies whose ISM is
being disproportionally affected by star formation properties 
(Leitherer 1997).  These galaxies are extremely important for understanding a 
host of astrophysical problems.  It is estimated that at least 25\% of all
star formation in the local universe is occurring in starburst galaxies 
(e.g. Gallego et al. 1995), with potentially a significantly larger 
fraction in the past (Lilly et al. 1996; Madau et al. 1996).  
Likewise, nearby starbursts have some of the highest surface brightnesses 
of any  galaxies in the universe, and have morphological and spectroscopic 
properties similar to high redshift galaxies (Gallagher et al. 1989; Cowie 
et al. 1995; Giavalisco et al. 1996; Hibbard \& Vacca 1997; Heckman et al. 
1998; Meurer, Heckman \& Calzetti 1999).  
It is possible that a significant fraction of all old stars were
created in starburst like events.  Detailed studies of nearby 
starbursts are therefore important for understanding the interactions 
between the ISM of a galaxy and young stars,
as well as for understanding the processes of galaxy evolution.
 
   There are, however, several very basic questions about starburst 
galaxies that are just beginning to be answered.  It seems likely that a 
fraction of all starburst galaxies are caused by the interaction 
and possible mergers of two galaxies (Schweizer 1987; Jog \& Das 1992).  
These starbursts usually have obvious companions producing an interaction 
in part responsible for triggering the starburst.  However, some starbursts
appear relatively isolated, and are either merger remnants of two
galaxies, or have a starburst mechanism not related to dynamical effects
from other galaxies. Starbursts can be triggered
from a self-contained internal process, such as from energy ejecta from 
super-winds and SNe explosions (Heckman, Armus \& Miley 1990).  Alternatively,
starbursts in the nuclei of galaxies can be triggered by bar-instabilities
(Shlosman, Begelman \& Frank 1990).  If there are different 
methods of triggering a  starburst, which seems likely, then are there also
differences in the morphological and physical properties of galaxies that 
have different starburst creation scenarios? 

  Some UV bright starburst galaxies are also bright in far-infrared 
wavelengths, revealing the presence of large amounts of dust (e.g.
Hunter et al. 1989).  Since 
UV light is strongly absorbed by dust, on the surface is an apparent 
contradiction which can be solved if the dust in galaxies exists in 
a 'picket-fence' geometry (Calzetti 1997). 
There are nearby examples of starburst galaxies undergoing vigorous
star formation, which have high far-IR fluxes, but little to no ultraviolet
(UV) flux (Sanders \& Mirabel 1996).   These ultra-luminous infrared galaxies 
(ULIGS) have become increasing important objects for understanding high 
red-shift sub-mm galaxies (e.g. Barger et al. 1999).  The number of known 
sub-mm  sources in areas such as the Hubble Deep Field are few, and it is 
difficult to determine if any correspond to UV- bright star forming systems 
at high redshifts ($>2$).  By investigating the spectral
energy distributions of nearby starbursts, we can get a better idea
of what to expect in terms of sub-mm source counts if most high redshift
galaxies resemble nearby starbursts.

  Previous  morphological studies of UV-bright starbursts have included 
the characterization of luminosity functions of
super star clusters (SSCs) (e.g.  Meurer et al. 1995; Maoz et al. 1996; 
Johnson et al. 1999) to test the idea that these objects are
pre-evolved globular clusters.  Meurer et al. (1995) investigated
the UV morphology using Faint Object Camera (FOC) images for nine galaxies, 
finding a highly
irregular appearance of the starburst region of each galaxy, and
most of the SSCs in the core of the starburst. 
Maoz et al. (1996) used FOC UV images, taken
with the Hubble Space Telescope, to investigate star formation
properties in ringed starburst galaxies.  Like Meurer et al. (1995), 
Maoz et al. find a large percentage of the UV light in starburst
galaxies, up to 50\%, to be in star clusters.  That is, a significant
portion of star formation seen in the UV-visible spectral region
occurs in dense clusters distributed throughout the starburst.   

In this paper, we investigate the large-scale physical morphologies
of UV bright star forming zones in nearby starbursts, and explore how 
the UV morphology compares to the optical appearance.  This will help 
determine the morphological evolution
of these objects and will reveal any implications for high redshift 
galaxies.   We also investigate how the 
spectral energy distributions of starbursts relate to their morphological 
appearance.

  For this study, We have obtained archival FOC/WFPC2 Hubble Space Telescope
ultraviolet imaging data, coupled with R, B \& H$\alpha$ images taken with
the 3.5m WIYN\footnote{The WIYN Observatory is a
joint facility of the University of Wisconsin-Madison, Indiana
University, Yale University, and the National Optical Astronomy
Observatories.} telescopes for the 
starburst galaxies: NGC 3310, NGC 3351, NGC 3690, NGC 3991, NGC 4861, 
and NGC 7673.  These galaxies represent a range of starburst 
galaxy host morphologies at luminosities near those of typical giant
spirals.  All of these galaxies, with the
exception of NGC 4861, are both bright in the UV and have a high 
far-infrared flux as measured by  IRAS.  We
examine the morphologies of these galaxies, including the comparison
between the H$\alpha$ and UV emission from star forming regions.  H II regions 
mark the locations where a mixture of stars with ages of less than about 3 Myr
are present; if the region is sufficiently dusty, then the
UV emission can be absorbed.  It is therefore useful to determine if 
there are any morphological
differences at various wavelengths, and what they imply for the 
physics of starbursts.
This is also important for understanding if nearby starbursts
are similar to high redshift ones, since many high redshift galaxies are
most easily observed in their UV rest frame.

   We find for our small sample that the starburst morphologies 
remain similar at different wavelengths, although some differences 
exist between the UV and H$\alpha$ morphologies.  We attribute these to  
aging and dust absorption effects within star-forming complexes.  The 
similarity between morphologies is a confirmation of the picket fence model 
of dust distributions in starbursts. The high FIR fluxes from these galaxies 
are the result of dust extinction occurring at places other than the 
UV-bright star formation sites.

   We also show that the slope of the spectral energy
distribution in some cases correlates with morphological
features of a starburst.  Our best example is NGC 3690, which has a disturbed
morphology, the highest dust content, and the largest FIR SED slope.  We 
conclude, based on similar panchromatic
studies of high redshift galaxies, that nearby starbursts, and high redshift
star forming galaxies have similar morphological properties.

In Section 2 we describe the motivation for this study; Section 3 gives
our observations and the data used in this study.
In Section 4 we explain in detail properties of the various starbursts
and discuss the morphological
differences between the wavelengths used in 
this study.  Section 5 is a description of our results, while Section 6 is an
attempt to explain the observed morphologies by invoking various physical
scenarios.   Finally in Sections 7 and 8 we examine how the morphologies
are related to both the spectral energy distributions and how these
galaxies are related to star forming systems at high redshift.
Section 9 is a summary of our conclusions.

\section{Physical Motivation}

   The study of nearby starburst galaxies is extremely important for 
understanding how star formation occurs - a basic question in contemporary
astronomy. Although active star formation in most galaxies is  
quiescent, localized, and may differ from the
intense star formation found in starburst galaxies, the basic processes
should be similar on the microscopic scale.  Studies of 
galaxies where star formation is dominating the optical appearance may 
reveal general features of star formation unobtainable from more quiescent 
galaxies.  

   The modeling of how star formation occurs on large scales is complicated by
many factors, the least of which is how massive individual stars
collapse to begin the nuclear fusion process.  Other factors
which will affect the panchromatic morphology of a star bursting region are: 
dust, photon ionization of the surrounding ISM, stellar-winds, SN energy 
ejecta, and ionizing photon leakage from other star forming regions.  
Combined with these on-going directly measurable features (in principle)
are the following: history and age of the starburst, the star-formation
rate of the burst (continuous as opposed to instantaneous star 
formation), as well as the details of the star formation process such
as the initial mass function, initial metallicity, spatial distributions,
and the evolutionary history of the new stars.   Synthesis models have
recently made efforts to reproduce photometric parameters of starbursts
as a function of time (e.g. Leitherer \& Heckman 1995; Leitherer et al. 1999).
These models consider all of the above parameters to derive measurable
quantities such as colors and luminosities throughout the history of
the burst.

   By combining morphological information with physical data known
for these galaxies, we can determine if the panchromatic morphological 
differences in these starbursts are due to an active force, such as the
effects of stellar ejecta, aging effects of the cluster, or alternatively
from dust obscuration effects.  Likewise, we want to understand
how the morphologies of UV-bright starbursts reflect the star formation
process and/or the dynamical state of the system.

\section{Observations and Data}

   This paper uses images taken with the WIYN 3.5m telescope in broadband
Harris B, R and narrow-band H$\alpha$, filters and HST WFPC2
and Faint Object Camera (FOC) archival 
ultraviolet images.  Table 1 lists the galaxies in our
sample with the filters used in the imaging.    The sample
consists of six galaxies of various starburst morphologies that were picked
based on the following criteria.  We used the UV spectral
catalog of Kinney et al. (1993) and the IRAS bright galaxy catalog (Soifer et 
al. 1989) to determine which starbursts
have both a high flux in the far-UV and a high far infrared flux as seen
by IRAS.   From this sample we then narrowed our candidates to the ones
that were imaged in the UV at $\lambda$ $< 0.3 \micron$ by the Hubble Space 
Telescope.
Each of the starbursts are relatively nearby and for the most part can be 
considered benchmark starbursts.   A detailed description and background of 
the individual starbursts will be given in Section 4.

  The WIYN optical observations were carried out between the nights of 
March 21 and March 24, 1998.   The WIYN CCD images were obtained with a 
2048$^{2}$-pixel 
thinned S2kB device, producing images with a scale of 0.2 arcsec per pixel.  
The field of view of each image is 6.8 x 6.8 arcmin$^{2}$.  Seeing condition 
for most of the images are $<1''$, with a few exceptions.   While the seeing 
was good, the conditions were not photometric.  The analysis presented in 
this paper is therefore based on morphology and relative photometry.

   A major part of this study is to compare the H$\alpha$ structures of
starbursts to those in the UV, hence adequate angular resolution 
UV images are necessary for this work.  Most previous UV imaging studies
of galaxies (e.g. Buat et al. 1987; Donas et al. 1987; Chen et al. 1992)
investigated mainly nominal normal galaxies and were limited in
resolution to a few arc-minutes.   The Ultraviolet Imaging Telescope (UIT)
has in particular produced a unique archive of moderate angular resolution
far ultraviolet images of galaxies, but observed only a few
starbursts (e.g. NGC 3310 (Smith et al. 1996)).  The superb resolution
offered by HST offers the additional advantage of allowing us to study 
the fine details of the starburst morphology.

  Four of the six galaxies in our sample: NGC 3310, NGC 3690, NGC 3991, 
NGC 4861, have UV images taken with the
FOC in the f/96 mode.  This gives a 14'' field of view with a resolution
of 0.014'' per pixel.  The filter used was the F220W, giving an
effective wavelength of 230 nm for these observations.  Figure 1 shows
a WIYN image of each galaxy, and the location of the UV images are marked by
a dark circle.  Usually the
UV images cover only a small part of the star forming area of these
galaxies.   The exposure
times for these galaxies were: 200s, 900s, 400s, and 1000s.  The basic
parameters for these observations are shown in Table 1.  The UV measurements 
for the other two galaxies, NGC 7673 and NGC 3351, are Wide Field Planetary 
Camera-2 (WFPC2) images  with fields of view 2.5' on a side with a 
resolution of 0.10''.  The WFPC2 filters
used for these galaxies are F218W and F255W with exposure times of
600s and 800s, respectively.   Photometry is
performed on these UV images of our sample by using the STMAG system, where
m = -2.5 x log$_{10}$ {\it f$_\lambda$} - 21.10.  We use an aperture that
covers the entire UV image, this is usually well defined in the cases 
of these starbursts, where most of the flux comes from centralized well-defined
star clusters. 

  For the spectral energy distributions, we use the International Ultraviolet
Explorer (IUE) UV fluxes.  Infrared Astronomical Satellite (IRAS) data are 
used for the Far-Infrared, and various sources, including the RC3 
(de Vaucouleurs et al. 1991) are used to obtain the optical and near infrared 
data. Without a homogeneous data set, it is nearly impossible to obtain 
fluxes within the same aperture over a wide range of wavelengths.
This is the case here, although it was tried to match aperture as
closely as possible for the UV and optical. The plots 
used to show the SEDs (Figure 8,9 \& 10) have substantial uncertainties in 
their aperture corrections, and in no other
part of this paper are different data sets compared with each other.

\section{Individual Galaxies}

\subsection{NGC 3310}

   NGC 3310 is one of the best examples of a local UV-bright starburst of
moderately high luminosity whose
unusual outer structure is probably the result of a recent merger with a 
smaller galaxy (Balick \& Heckman 1981; Mulder \& van Driel 1996).
NGC 3310 is also extremely bright in far-infrared emission (e.g. Braine 
et al. 1993), X-rays (Zezas et al. 1998), and was one of the first known 
FUV bright starbursts (Code \& Welch 1982).

   In the optical NGC 3310 displays a striking 'bow and arrow' 
appearance in its outer parts.  This feature is one of the better studied
morphological peculiarities in any starburst galaxy (Walker \& Chincarini 
1967; Balick \& Heckman 1981;  Bertola \& Sharp 1984; Mulder et al. 1995).    
It was once thought that this bow part was a spiral arm 
(Walker \& Chincarini 1967), but has been shown as part of a shell 
or ripple pattern, possibly caused by the accretion of a low mass companion
galaxy (e.g. Balick \& Heckman 1981; Schweizer \& Seitzer 1988; Smith et al.
1996).

  NGC 3310 has an inner, symmetric spiral pattern and a 
significant amount of star formation tracing the spiral arms.  A 
large portion of the star formation, based on H$\alpha$ images, is in a 
central ring surrounding an off-center nucleus (Figure 2).  Some 
star formation is occurring within the ring, but the nucleus
is either dominated by an older population, or the nucleus
is dusty (compare UV and R band images in Figure 2).

   The FOC UV image of NGC 3310 shows a similar morphology to the 
H$\alpha$ image;  there is a ring of hot stars surrounding the central 
parts of the cluster.  However, when comparing directly the UV, R and 
H$\alpha$ morphologies, there are some obvious differences (Figure 2). There 
is a double ring structure in the core of the galaxy, an inner one 
composed of gas (H$\alpha$ emission), and the outer one composed of slightly
older stars.   The inner H$\alpha$ ring is immediately interior to the redder
outer ring, and is off-set from the center of the galaxy.  However, this 
outer ring is not dominated by H$\alpha$ emission as
much as the arms leading away from the center are.  Additionally there are
stellar regions in the center which have little or no H$\alpha$
emission; likewise there are several regions near the western
part of the center that are dominated by H$\alpha$ flux, with no continuum
R light.   

   By comparing the H$\alpha$ and UV image for this galaxy (Figure 2) 
several regions at the upper left are found to be brighter in 
H$\alpha$ than in the UV.   These are also areas with considerable H$\alpha$ 
flux 
but little UV flux in comparison to the other bright UV regions.  There is 
an asymmetry in the difference, with the 
north-west region having less UV flux than the south-west region of the 
central ring around the galaxy.   We conclude that this starburst is
probably a result of star formation induced by a bar instability, an issue
we will discuss more in Section 7.1.  The color map of this galaxy does not
show large variations, or the clumpy appearance of dust.  

\subsection{NGC 3351}

 NGC 3351 appears to be symmetric in both its outer and inner portions, and 
with the exception of the nuclear starburst, appears quite
regular with none of the tidal features that are observed in many of the 
other galaxies
in our sample (Figure 3).  The classical morphological 
description of this galaxy is a ringed-barred Sb, while most UV-bright 
starbursts have classically been defined as either irregular or peculiar
galaxies.   The regular morphology of NGC 3351 suggests that this galaxy's
starburst has a different origin from the others in our sample.
NGC 3351 has a prominent bar, and the dynamical influence of the bar is a 
likely source of this starburst (Shlosman et al. 1990).

  Previous work on NGC 3351 (e.g. Alloin \& Nieto
1982; Moaz et al. 1996) found three H$\alpha$ bright H II regions surround 
the nucleus.  
Kinney et al. (1993) also find highly ionized species of heavy elements 
indicating the presence of intense radiation, consistent with the central
starburst interpretation.  

  We find four circum-nuclear H II complexes in our H$\alpha$ images, and also
see several UV star clusters that have neither corresponding H II regions, 
nor a significant H$\alpha$ flux.  One interpretation for the existence of 
the star burst in 
NGC 3351 is that the gas is being driven into the ring around the nucleus 
creating the starburst in the inner Lindblad resonance 
(Alloin \& Nieto 1982).    The lifetimes of intensely star-forming nuclear
rings in giant disk galaxies are not very well-known, and this class
of UV-bright starburst may be fundamentally different from those produced
by interactions.

\subsection{NGC 3690}

    NGC 3690 is  the most disturbed 
galaxy in our sample, and is morphologically classified as 'peculiar' (see
Figure 4).  It is the only galaxy in our sample that is interacting  
strongly with another galaxy, in this case a probable merger with
IC 694 (Gehrz, Sramek \& Weedman 1983).

  HI mapping of this galaxy finds no disk structure, and its optical 
morphology can be accounted for primarily by bright star forming regions 
 (Nordgren et al. 
1997).   Of our entire sample, NGC 3690 has the largest 
difference between ultraviolet and H$\alpha$ morphologies in the inner portions
of the starburst.    It has both H II regions and UV star clusters without 
the corresponding presence of the other.    

   There are four main star forming regions in the
inner parts of this galaxy pair, called nuclei A, B, C, and C$^{\prime}$ 
following Sargent and Scoville (1991).  These clumps are thought to be the 
main source of the prodigious far infrared luminosity of this system. The 
mid-infrared emission (10 $\mu$m) from these sources is extended (Miles et 
al. 1996), and compared with 3.4 $\mu$m emission, the source C is bluest, 
followed by B, with A and C$^{\prime}$ the reddest. NGC 3690 is one of the 
brightest galaxies 
in the local universe in X-rays, with a luminosity L$_{x}\approx 10^{42}$ 
erg s$^{-1}$ and an X-ray spectrum consistent with a super-wind 
(Zezas et al. 1998).  

   Given the very peculiar morphology, the starburst of this galaxy is
a result of the on-going galaxy merger with IC 694. The color 
map of this far infrared bright galaxy 
also shows that a significant amount of dust is present (Figure 3).  Not,
surprisingly, this galaxy has the largest ratio of FIR to starburst
UV flux (see Section 7).

\subsection{NGC 3991}

    Cataloged as a peculiar galaxy by Arp (1966), NGC 3991 has also been 
classified as a Magellanic irregular (see Figure 5).   NGC 3991 has an 
optical spectrum similar to an H II galaxy (Keel et al. 1985; 
Kennicutt 1992).    NGC 3991 also has low metallicity H II regions 
(Arnault et al. 1988), and from our images does not appear to contain 
an extended shell, or outer portion distinct from its inner starburst 
area.  However, it is generally believed that the starburst in this 
galaxy was triggered by an interaction with the neighboring
galaxies NGC 3994 and NGC 3995 (Keel et al. 1985).   A detailed optical and 
spectroscopic study of NGC 3991 (Hecquet et al. 1995) found several
separate knots in NGC 3991 each with extreme star formation 
and sizes of around 300 pc.  A comparison of the colors of these knots
indicates that star formation is concentrated within and occurring
rapidly in these clumps (Hecquet et al. 1995).

 NGC 3991 also has a high radio and X-ray flux (Seaquist \& Bell 1968; 
Fabbiano et al. 1982). The UV spectrum of this galaxy contains features 
associated with recently formed stars such as Ly $\alpha$ and He II emission
(Kinney et al. 1993).
Although detected by IRAS, NGC 3991 is not classified as an IRAS bright
galaxy (Soifer et al. 1989).  Therefore,  the morphology of NGC 3991
is probably not dominated by dust.  We verify this, by finding the
UV morphology is similar to the optical R band.  However,
the southern H II regions associated with a relatively
bright UV cluster are faint in H$\alpha$ as compared with the northern
star-forming clumps.  

\subsection{NGC 4861}

  NGC 4861 (Arp 266) has no regular
structure, or obvious outer stellar envelope, and appears to be a strand of 
H II regions. NGC 4861 is classified as a Magellanic irregular 
(Sandage \& Tammann 1981) (see Figures 1 \& 6).
The northern part of this galaxy has occasionally been considered a separate
system, IC 3961.   The southern area, appearing as a bright knot is sometimes
referred to as NGC 4861.  The knot is usually interpreted as an H II region
or OB association (Huchra 1977a).  From HI maps, this galaxy is an edge-on
and rotating disk system (Wilcots, private communication.)

 NGC 4861 has been studied as an example of a compact blue galaxy 
(Thuan \& Martin 1981), having a high UV continuum and strong absorption
features, indicating the presence of young O and B stars (Kinney et al. 1993;
Calzetti 1997). Radio continuum work on NGC 4861 finds a deficiency of 
non-thermal emission (Sramek \& Weedman 1986; Collision et al. 1994), possibly 
indicating a lack of supernova remnants.

   The R and B WIYN images clearly show the southern knot to
be very compact and unresolved in 1'' seeing, while the northern part of 
this galaxy is resolved into stars or stellar clusters.  The FOC image of NGC 
4861 resolves the southern knot into six stellar condensations,
although these probably do not completely account for its bright optical
appearance.  The OB clusters are generally on the outskirts of the
southern knot and it is likely that the ionizing photons from these 
clusters are the source of ionization for the surrounding H II region.  The 
ionized gas,  delineated by the H$\alpha$ flux in and surrounding the knot is 
more diffuse and has a larger extent than the stellar light; this behavior
is typical of giant H II regions.  The 
morphological appearance of the southern knot is largely the  
result of this ionized gas.

  The FOC image includes also a part of the northern part of the starburst, 
where only one small UV source (OB association or cluster) can be seen, 
with a considerable amount of
diffuse UV light caused by unresolved OB stars.
For the most part the morphology of NGC 4861 is 'simpler' then the
other galaxies in our sample, since there are very few major H II regions
resolved in our H$\alpha$ image.  From what we can determine there
is no significant difference between the UV and optical morphology of
this edge-on starburst.  

\subsection{NGC 7673}

    NGC 7673 is a nearby, luminous starburst galaxy with an inner asymmetric
spiral structure and an outer structure showing evidence of
ripples.  This ripple structure of NGC 7673 is similar to 
that in NGC 3310, possibly caused by a merger or interaction with
another galaxy (see Homeier \& Gallagher 1999).  NGC 7673 also contains an 
extensive array of clumpy star forming regions throughout its disk (Casini \&
Heidmann 1976). The starburst in this galaxy is 
occurring in the inner portions of the galaxy within huge clumps
embedded within an abnormal spiral 
pattern (Huchra 1977b; Gallagher et al. 1999).   There is also an 
extended, somewhat disturbed HI disk (Nordgren et al. 1997).   The most 
likely scenario for the creation of the starburst in
NGC 7673 is either through a minor merger, or more likely from
an interaction with a nearby galaxy, NGC 7677.

 Although a massive starburst is occurring in the disk of NGC 7673, the
H$\alpha$-line kinematics are relatively quiescent, with a low velocity 
dispersion (Duflot-Augarde \& Alloin 1982; Homeier \& Gallagher 1999) 
consistent with a rotating disk.  Figure 7 shows the UV/H$\alpha$ and
optical images for this galaxy.  We find that NGC 7673 has similar
structures in its UV and H$\alpha$ images, with most H II regions following
the locations of the bright UV star clusters.  A more detailed discussion
of star formation patterns in NGC 7673 will be presented by Gallagher 
et al. (1999).

\section{Color Maps and Morphological Band Differences}

   A color map of a galaxy, showing the relative proportion of light
from one band to another, is useful for determining the presence of
several features in a starburst.  First, by examining these maps,
it can become clear if dust screens are strongly affecting the light 
distribution.  This is typically revealed by narrow, or localized features 
(e.g. dust lanes) where the redder band dominates the bluer one.  These 
features will generally 
increase in prominence at shorter wavelengths, until the
background becomes too faint in the UV.  Likewise, a color map will also 
reveal the locations 
of the bluest, and hence the least obscured massive stars.  In this study,
we display the color maps for the starburst regions of interest to demonstrate
the existence or lack thereof, of dust and the locations of star formation
activity.

   From our multiple wavelength images, we can also study the
changes in morphology from long to short wavelengths, between the
wavelengths of 220 nm to 660 nm.  We find that although the host galaxy
for the starburst essentially disappears in the UV image, the star bursting
region's large scale morphology is usually similar in all bands.   

 Of course, when comparing the rest frame morphologies of galaxies in the FUV
with visible wavelengths, differences do exist (O'Connell \& Marcum 1996), 
which can be attributed to the sampling of different stellar populations.  
What is the behavior of the panchromatic morphology for distant star-forming
galaxies? When examining high redshift galaxies, the structure will also change
due to cosmological surface brightness dimming.  When observing galaxies
at high redshifts, such as those in the Hubble Deep Field the 
brightest portions of objects are seen; these are likely to be the most 
active sites of star formation.  The intensity of star-formation in some 
distant galaxies is much greater than local starbursts, by up to a factor of 
four (Weedman et al. 1998).  While the total intensity and rate of
star-formation is different, other macroscopic morphological features
are similar.    No large morphological changes are 
observed between the morphologies of Hubble Deep Field galaxies in the NICMOS 
infrared 
images (rest frame optical) and in the WFPC2 visible bands (rest frame UV) 
(Dickinson et al. 2000).  Either these images are too shallow to find fainter 
host galaxy-like components or these galaxies are dominated by stars emitting 
in the UV.  We discuss other implications of studies of nearby starbursts for 
high-redshift galaxies in a separate section.

\section{Physics and Geometry}

  Although our sample of UV and H$\alpha$ images is small, we can 
still draw useful conclusions about the ways in which multi-wavelength 
images sample different aspects of the distribution of stars and gas in 
these starbursts.  The H$\alpha$ images reveal the presence of ionized
gas and the locations of the most recent star formation events ($<$ 10 Myr), 
while the UV star clusters reveal the presence of young stars recently
formed over the last 50 - 100 Myrs (O'Connell 1997).  Since young stars 
are formed out of gas, it might be expected that
the light from UV clusters would trace that of H$\alpha$ gas ionized by the
high energy UV photons.  Yet, we do see luminous H II regions without FUV
sources and FUV regions without H$\alpha$.

    Possible explanations for these 
differences include: dust absorption of the UV light, aging effects, 
UV photon leakage into areas of previously neutral hydrogen gas, 
as well as intense winds from young stars and supernova that could 
rearrange interstellar gas from which OB stars formed.  
We will discuss below each possibility in detail and arguments for and 
against explanations for describing the different morphologies.

   The fact that we often do not see a major difference between the UV and
optical morphologies tells us about the geometry of the stars and dust
in these starbursts.  If there were significant interstellar clouds along 
the line of sight towards a starburst, then these would introduce strongly 
wavelength dependent morphologies.  Since we do not see
this effect, we conclude that for the most part the structure of
these starbursts reflects the localized concentration of dust and young
stars in clumps or relatively thin sheets.

\section{Effects from Dust}

   Dust is one of the key components of starburst galaxies at both high
and low redshifts (e.g Calzetti, Kinney \& Storichi-Bergmann 1996; 
Calzetti 1997; Calzetti 1999).  The presence
of dust is a critical aspect of star formation, and it is not surprising that
a very significant amount is found associated with it
in all environments.  Besides
absorbing light, dust also re-radiates the bulk of its energy
at temperatures around 30-60 K,
thermally producing a high far-infrared
flux from starburst galaxies; effectively transferring flux from
short UV wavelengths to longer far infrared ones.    
If a starburst (or any object) were to have a high UV and FIR flux, this
could be construed as a potential paradox if we believe that the dust
is absorbing a significant amount of the UV light.

   An important factor in resolving this apparent
paradox is how the dust in starbursts is distributed.  The common assumption
is that some of the dust is distributed in an inhomogeneous 
foreground screen which has been verified by both observations and numerical
modeling (e.g. Calzetti et al. 1994; Meurer et al. 1995; Calzetti et al. 1996; 
Calzetti 1997; Gordon, Calzetti \& Witt 1997).   This 'picket fence'
distribution of dust can explain several features of the starbursts
in our sample. 

   Absorption of light by dust will affect the stellar continuum from the UV
to the IR, but its impact on morphologies is larger at shorter wavelength
images.  For our sample  we 
can directly determine the obscuration and dust distributions  
by examining both the morphology and integrated fluxes of the UV images as 
compared with the optical.  
If dust is directly affecting each star bursting region the same way with
similar geometries, and if the composition of the dust is the same, then we 
should expect the wavelength dependence of obscuration for each region to be 
similar.  Any deviations between regions would indicate that the properties of
the dust are varying, or more likely that the dust and stars are distributed 
differently in different star bursting regions.    We find that for the most
part, the morphologies of the UV galaxies are almost identical to the
optical morphologies.   Foreground dust screens are therefore not major
absorbers of UV light in most of our sample.  Counter examples are UV-dim
H II regions, where dust screens are evidently blocking the UV light.

    This however, does not imply that dust is an insignificant component
of the SED of these galaxies.  We know dust
is present from the high far infrared fluxes of these galaxies.  A more
likely explanation is that the dust is being either destroyed at
locations of intense star formation (Draine
\& Salpeter 1979; Calzetti et al. 1996) or is being moved to other parts
of the galaxy from energy ejecta from stellar winds and supernova (Heckman,
Armus \& Miley 1990; Calzetti et al. 1996).    The galaxies where we can see 
distinct patchy dust absorption all show these features away from the 
areas of the star clusters and H II regions.   The dust is probably being
removed from the star forming regions into the more hospitable
 environments away from the starburst clumps.  Dust in areas away from the
star clusters allows a starburst to be both a high UV and FIR emitter, 
and to have a largely unextincted UV morphology.   Its worth mentioning that
some of the FIR emission might arise from young clusters completely surrounded 
in dust, and invisible in any of our images.  This component would only
make up a small fraction of the FIR emission however, and most of it
must be coming from these inter cluster regions.

\section{UV/Ionizing Photon Leakage}

    Traditionally a much less studied effect on the panchromatic morphology
of galaxies than dust or winds is the effects caused by Lyman continuum
photon leakage from active H II regions to ionize gas in otherwise quiescent 
pockets of gas.   This leakage of photons from star forming regions has been 
documented for several galaxies (e.g. Ferguson et al. 1996).  
The leakage of photons from H II regions has long been a popular explanation
for the significant amount of diffuse ionized gas (DIG) found in nearly
all types of galaxies with star formation, including dwarfs (e.g. Hunter, 
Hawley \& Gallagher 1993; Martin 1998),
starbursts (e.g. Calzetti et al. 1999a), irregular (e.g. Hoopes et al. 1996; 
Otte \& Dettmar 1999) and in late-type spirals (e.g. Wang et al. 1997), as 
well as the Milky Way (Reynolds 1985).

  The study here is not designed or suited to study the diffuse
interstellar gas present in these starbursts, expect to note
that a diffuse component does exist, especially in  NGC 4861 and NGC 3690.    
We are interested in trying to explain an observed H II region with out
a corresponding UV star cluster, by this method of
ionization.  These H II regions appear roughly 
symmetrical, but we do not see a corresponding UV star cluster in the
region. If ionizing photons were responsible for producing this H II region,
they would have to do so in the observed symmetrical pattern, an event
that seems unlikely.  The diffuse H$\alpha$ gas detected away 
from the regions of star formation are probably ionized by photon leakage
from UV star clusters.  However, we discount photon leakage from
neighboring star clusters as a possibility for 
explaining an H II region without a corresponding UV star cluster.
The likely cause is either from dust absorption, or aging effects.

\section{Kinematic effects from Winds and Supernova}

   A very significant physical component to the evolution of any star forming
region, particularly regions with high mass star formation, are the
kinematic effects from stellar winds and SNe explosions.   As mechanical
energy from SNe and strong stellar winds from both OB and WR stars act on
the gas in a star forming region, it will input a certain amount of
kinetic energy.  This energy will interact with the ISM from which the
star form.  When this occurs, it is likely that a fraction of the gas
surrounding the newly formed stars will be expelled away from the area of
star formation.  Likewise, the effects of winds and SN combined to
create super-winds can also be responsible for triggering star formation, as in
the well know case of M82 (e.g. Satyapal et al. 1997).     While most of
our starburst sample seem to be triggered by an interaction or bar instability,
the resulting winds maybe partially responsible for the starburst
morphology.  This would require kinematic data to prove.
 
  We do not claim that any of our sample starbursts are triggered solely
by a wind, but the
fact that we can see clearly the UV star clusters is due in part to
stellar winds and energy ejecta clearing away dust grains.  Mass outflows from
young stars is 
an adequate method for removing or destroying the dust near the bright star 
clusters (Heckman et al. 1990).  The morphological signature for a starburst 
extremely affected by energy ejecta
would be a very bright young UV star cluster with no corresponding H II region.

  We see this in the case of NGC 3351, which is also the only galaxy
in our sample that does not have any tidal features.   The other UV
star clusters appear similar in the UV and optical, and
therefore a significant amount UV light is not being absorbed. The dust 
that was originally associated with the young star clusters is being either 
destroyed, or more likely,
based on a high FIR flux, moved to other areas of the starburst where the 
star formation is more quiescent.    This scenario allow the UV star clusters
to remain bright in the UV, with the galaxy still emitting a high FIR
flux due to the absorption of photons from the inter-cluster areas.  This
is confirmed by the measurement of a low temperature for  most of the dust 
in starbursts at around 20 - 25 K (Calzetti et al. 1999b).  This
low temperature for the dust implies it lies quite far away from the
site of the starburst.   Dust heavily associated with the star bursting regions
would have higher temperatures T $>$ 30 K.

\section{Spectral Energy Distributions and UV Morphology}

   The spectral energy distribution of a starburst tells us at what
wavelengths most of the energy of a starburst is coming from.  From Figures
8, 9 \& 10 it can see that for all the starbursts in our sample, except
NGC 4861, most of the energy is originating from the far infrared region.  
This is not surprising since these galaxies are known to be bright in
the FIR, and also starburst galaxies.  We know from 
previous work (e.g. Soifer et al. 1989) that
the FIR emission from a galaxy is largely due to the dust grains.  
A simple question to ask is how does the UV morphology correlate with the
spectral energy distribution of a starburst galaxy?

  One of the ways to characterize the FIR flux from dust is to compute
a star formation rate based on the 12$\mu$ to 100$\mu$ emission, particularly
by using the longer wavelengths.  This method assumes that most of the 
FIR emission from a galaxy is a direct result of energy re-radiated by the 
grains after being heated by UV radiation coming from nearly created stars.  
There are however, problems with this assumption (Calzetti et al. 1995) which
may lead to a faulty value for the star formation rate.

   To answer this we compare spectral energy distribution slopes with 
morphologies.   We define the spectral slope of the UV emission,
as measured by the IUE between the wavelengths of 1500 \AA  and 2700 \AA, 
as $\alpha$.  We further define the FIR emission slope between 25 and 60 
$\micron$ as $\omega$.  For both of these, the slopes are computed by taking
the difference in the fluxes, and dividing this number by the change in 
the wavelength.  Using these two spectral indexes (see
Table 4), we can get an idea of how morphology relates to the spectral
energy distribution.  The majority of spectral energy slopes in the
FIR are around 30x10$^{-28}$ W m$^{-2}$ Hz$^{-1}$ $\mu$$^{-1}$ (Table 4).  The 
log of the ratio between the FIR and UV slope is almost always between
1.1 and 1.5.

   Two galaxy in particular stand out as having unusual spectral
energy distributions.  One is the merger/interaction NGC 3690, which
has both a very high FIR slope ($\omega$), and a high FIR slope relative
to the UV slope ($\alpha$) (see Figure 11).  This galaxy is also the
most disturbed dynamically, and shows the greatest differences in
morphology between different wavelengths.  The morphological differences are 
probably due to dust, as shown in both the SED, and the UV morphology.  

   Another interesting case is NGC 4861, the only galaxy in our sample
which is not a bright IRAS source and is a low metallicity Magellanic system.
This galaxy is also the only one with a negative UV spectral slope.
This indicates that FUV photons are escaping the regions of
star formation.  If this is the
case, then compared with the other galaxies in our sample, NGC 4861 is less
dusty in the sense of having lower UV obscuration optical
depths.  This would also explain the lack
of a significant FIR flux coming from this galaxy.    Other galaxies may
 behave in a similar way, and by knowing the spectral
slopes of a galaxy it is possible to begin understanding its morphology.

\section{Morphological Features of UV Bright Starbursts}

  Our UV bright sample contain examples of morphological features such
as rings and star clumps that can illuminate the star formation process; 
we briefly discuss these here.  For a more detailed discussion of the 
relationship between ringed UV starbursts and star forming regions see 
Maoz et al. (1995). Two of our galaxies, NGC 3310 and NGC 3351 have a 
nuclear ring structure, with NGC 3310 being the more regular and symmetric 
of the two.  Maoz et al. (1995) conclude, as we do for our two ringed 
starbursts, that 
these starburst rings are not produced by an active nucleus, but are from
bar instabilities.  
  
     The prominent H$\alpha$ morphology of NGC 3310 in comparison to the
UV morphology (Figure 2), suggests that the starburst in this ring is
very young.  It seems likely that such a large configuration, 1 kpc across,
 could not have formed after this young starburst triggered, due to the
time scales involved, but that the starburst has 
occurred in this ring formation.  NGC 3310 is a barred galaxy, as is 
NGC 3351, and models of star formation predict that bars can drive gas 
into the inner Lindblad resonance to produce a starburst.  The 1 kpc ring in
NGC 3310, is the scale sized predicted for such a ringed
 star formation effect (e.g. Athanassoula 1992; Piner et al. 1995).  Previous 
observations have begun to verify dynamical effects of gas in-flows due
to bar instabilities (e.g. Regan, Vogel \& Teuben 1997; Reynaud
\& Downes 1997).   In light of its young age and size, it seems likely 
that the starburst in NGC 3310 was produced by a similar effect.
   The UV structure of NGC 3310 is also dominated by diffuse light, 
or unresolved star clusters.  This is especially true in comparison to the 
other galaxies in our sample.  The clusters in NGC 3310 have sizes that 
range in size from 10 pc for the super star clusters, down to unresolved 
clusters  that are $<1$pc.  These is rather large range in cluster sizes, 
and may be a further indication of their youth.

   The ring in NGC 3351 is not as symmetric as NGC 3310, resembling
more of a horseshoe than a ring (see Figure 3).  The size of
this ring is about 1/4 kpc, with most of the UV light output originating
in the star clusters, which have sizes smaller than the clusters in NGC 3310.
The largest UV clusters in NGC 3351 are only about 3 pc.  The H$\alpha$ and 
UV morphologies for the starburst in NGC 3351 also resemble each other, 
indicating that this starburst is probably older than the one in NGC 
3310.  This starburst was probably also produced by a bar instability. 

\section{Starbursts at High Redshift}

  Star forming galaxies at moderate-to-high redshifts, such as the compact 
blue galaxies, or CNELGs (Jangren et al. 1999), and the recently 
discovered Lyman-break galaxies are undergoing star formation, 
and are significant populations at these redshifts (Steidel et al. 1996; 
Lowenthal et al. 1997).

  Nearby starburst galaxies have similar properties to the these distant 
star forming galaxies (Gallagher et al. 1989; 
Cowie et al. 1995; Giavalisco et al. 1996; Hibbard \& Vacca 1997; Heckman 
et al. 1998; Meurer, Heckman \& Calzetti 1999).   Although the Lyman-break
galaxies are forming stars at a much faster rate than the galaxies in this
paper (e.g. Weedman et al. 1998), both types are examples of starburst 
galaxies in a broad sense, and
are believed to be similar in terms of both stellar populations, and 
structure.   Local and high-redshift starbursts also contain a similar
star formation rate per unit area; although the distant galaxies can have
total star formation rates up to 10 times as high.

   This consensus is based on large scale characteristics,
such as star formation rates, magnitudes, and apparent morphologies, but
there may be more deeper problems with associating nearby starbursts
with high redshift galaxies, particularity if there are any metallicity
differences (see Leitherer 1997).  Measurements of the metallicities for
these Lyman-break galaxies is still in its infancy, but metallicities of
damped Lyman $\alpha$ systems are significantly more metal poor than nearby
starbursts (Pettini et al. 1997).  Measuring metallicities in distant
galaxies is difficult, but some early measurements (e.g.
de Mello in prep.) find Lyman-break galaxies with metallicities 1/3 - 1/4 
solar. If there are other physical differences between local and distant 
starbursts, they can be found by comparing multi-wavelength photometric
and spectroscopic features.
    
  In terms of morphology, and star formation rates, nearby starburst galaxies
are similar to high redshift galaxies when artificially redshifted
(Hibbard \& Vacca 1997).  The spectral features of starbursts
also imply they have similar stellar populations as distant
galaxies (e.g. Heckman et al. 1998; Meurer, Heckman \& Calzetti 1999).   
The high-surface brightness, and
spectrum, dominated by early type O and B stars is physical evidence that
abundant star formation is occurring in these galaxies.    However,
it is nearly certain that some properties of high redshift galaxies
differ from nearby starbursts,  and these features may have an effect
on the star formation process.  It is also largely uncertain what the
dust content of these high redshift galaxies are, but methods
have been able to estimate it (e.g. Meurer, Heckman \& Calzetti 1999).
 
   Recently a set of high-redshift galaxies have been observed that are very 
bright in the sub-mm and radio wavelengths (e.g. Barger et al. 1999).
These galaxies, most of which were discovered with the SCUBA array at
the James Clerk Maxwell telescope, are referred to as
SCUBA sources.  These SCUBA sources have
spectral energy distributions implying a high amount of rest frame
far-infrared light, and possibly very little to no emission in the rest frame
ultraviolet, and are very red objects when observed in the near infrared
(Smail et al. 1999).  Extreme high-redshift sub-mm sources are therefore 
probably analogous 
to nearby ultra-luminous infrared galaxies (Meurer, Heckman \& Calzetti 1999),
rather than UV bright starbursts.  However, the classical starburst, M82, along
with our sample, have properties that suggest a possibly similarity with
the sub-mm galaxies.  The spectral energy distributions for
the galaxies in this paper (Figures 8-10) have distributions indicating
a high far-IR flux, as well as a bright UV flux.  We have however already
discussed how these two can be reconciled.    The question remains however
if the high redshift sub-mm sub-mm sources are a distinct population from
the Lyman-break galaxies, or if high-z galaxies can be both bright
in the rest frame UV and in the far-infrared such as the galaxies
present in this study.  Further optical identifications of sub-mm galaxies
at high redshift are necessary before this question can be answered.
   
   Furthermore, due to k-corrections, when viewing high redshift galaxies 
in optical wavelengths the light being sampled is originating from the rest 
frame ultraviolet light.  This rest frame UV flux and morphologies of high
redshift galaxies  are being used in a variety of programs which could
be biased if the assumptions about the UV light are incorrect or
misleading.  As shown in this paper, the UV morphology of nearby starbursts
is similar to the optical morphology.  If high red-shift galaxies
have properties similar to nearby UV bright starbursts, which has been 
backed up by many different observations (Gallagher et al. 1989; 
Cowie et al. 1995; Giavalisco et al. 1996; Hibbard \& Vacca 1997; Heckman 
et al. 1998; Meurer, Heckman \& Calzetti 1999), then optical images
of distant galaxies sampling the UV should be morphologically very similar to
their rest-frame optical images.  Morphological k-corrections are
therefore small on large scales, a conclusion backed up by HST NICMOS near 
infrared observations of the Hubble Deep Field (Dickinson et al. 2000).    
Although it can be argued that our sample is too small to make general trends 
for all starbursts, it is likely that similar morphological differences will 
appear in other nearby UV bright starbursts (see also Dey et al. 1999; Lilly
et al. 1999).

\section{Conclusions}

   By examining the panchromatic morphologies and spectral
energy distributions of six nearby UV bright starburst
galaxies we have come to the following major conclusions:\\
1.  The locations of H II regions and bright UV star clusters are 
generally the same.  We interpret exceptions as being
duster, and therefore often younger, H II complexes, or older UV clusters.\\
2.  The UV morphology of starburst regions in starbursting galaxies
is on large scales remarkably similar to the morphology in optical bands.
This is also true for galaxies at high redshift in the Hubble Deep Field;
further evidence that nearby starbursts morphologically resemble high redshift 
star forming galaxies. \\
3. The dust creating the high far-infrared flux for our sample is
not in general obscuring the bright UV star clusters, but must exist
in other more hospitable areas of the host galaxy, allowing
these starbursts to be both bright in the UV and FIR.\\
4. The spectral energy distributions slopes of starbursts can indicate
morphology in the sense that large FIR slopes correlate with disturbed
morphologies affected by dust.   Likewise, a negative UV slope reveals
small amounts of dust.

We are pleased to thank the crew at WIYN for the excellent operation 
of the telescope during our programs, and Stephan Jansen for his 
continuing support of astronomical computing at Wisconsin.
Support for this work was provided by NASA through grant number 
AR-07529.01 from the Space Telescope Science Institute (STScI), which is 
operated by the Association of Universities for Research in Astronomy, 
Inc., under NASA contract NAS5-26555.  CJC acknowledges the support and 
hospitality of STScI where part of this work was completed.

\clearpage

\begin{figure}
\plotfiddle{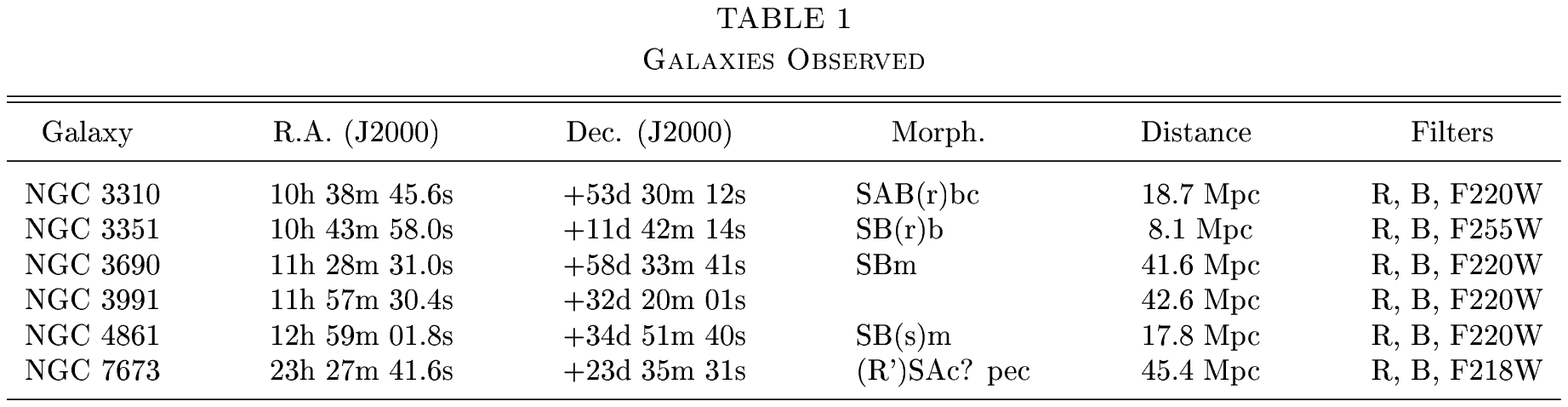}{6.0in}{0}{100}{100}{-310}{-170}
\end{figure}

\clearpage

\begin{figure}
\plotfiddle{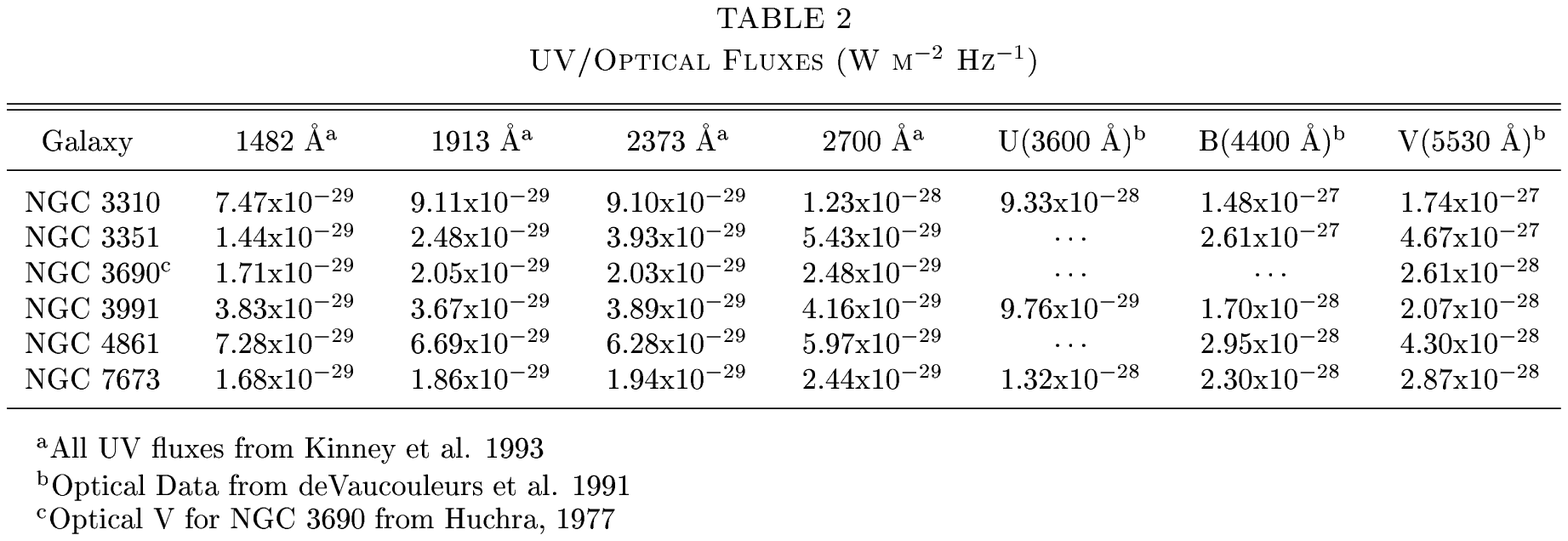}{6.0in}{0}{100}{100}{-310}{-170}
\end{figure}

\clearpage

\begin{figure}
\plotfiddle{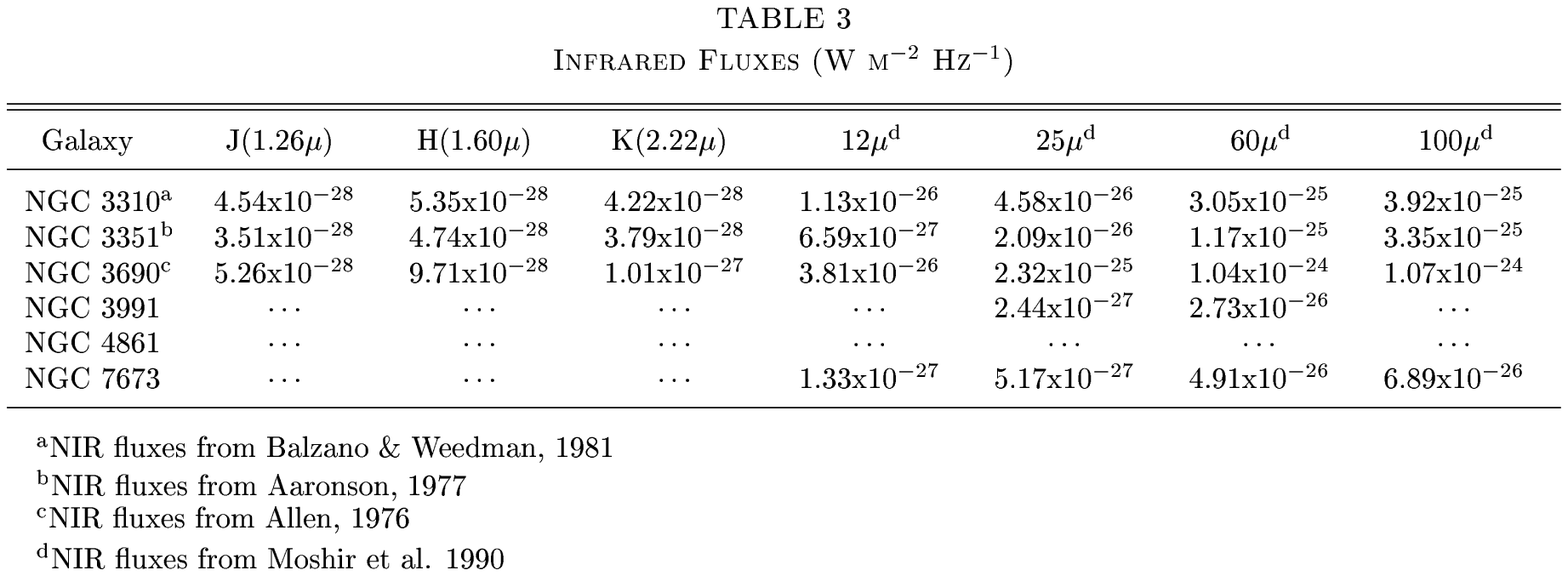}{6.0in}{0}{100}{100}{-310}{-170}
\end{figure}

\clearpage

\begin{figure}
\plotfiddle{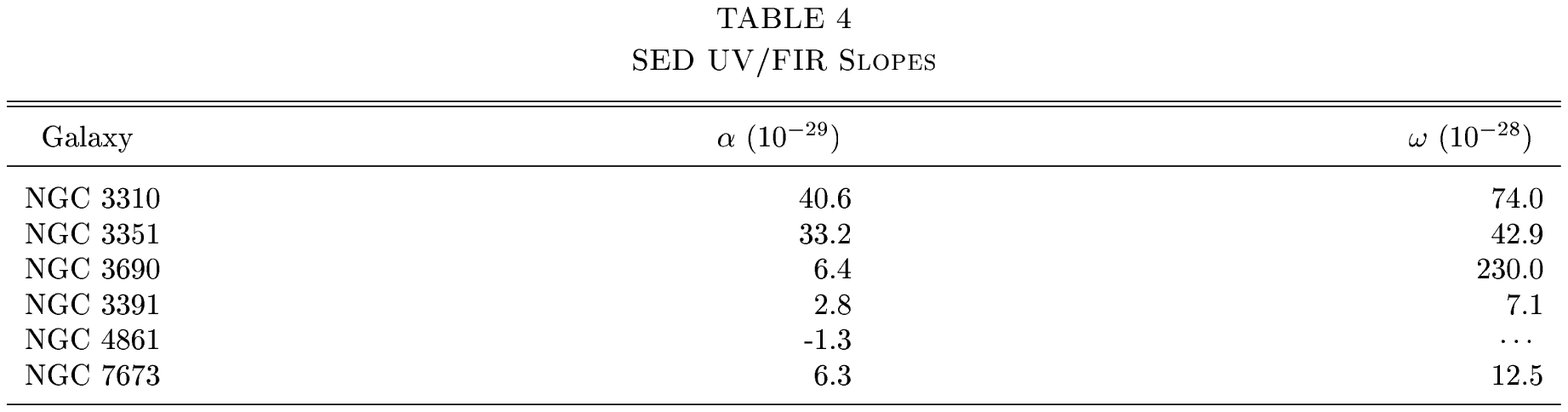}{6.0in}{0}{100}{100}{-310}{-170}
\end{figure}

\clearpage

\figcaption[]{Large scale image of each of the starburst in our sample.
The dark circle outlines the are observed in the UV images presented in
Figures 2 - 7.}

\figcaption[]{The panchromatic morphologies for NGC 3310.
Clockwise, the images are in the R band, B/R color map, FOC UV, and
the H$\alpha$ image.  The B/R map is oriented such that the brighter
portions are bluer and the darker are redder.}

\figcaption[]{Panchromatic morphologies for NGC 3351. See
Figure 2 for explanation of frames.}

\figcaption[]{Panchromatic morphologies for NGC 3690. See
Figure 2 for explanation of frames.}

\figcaption[]{Panchromatic morphologies for NGC 3991. See
Figure 2 for explanation of frames.}

\figcaption[]{Panchromatic morphologies for NGC 4861. See
Figure 2 for explanation of frames.}

\figcaption[]{Panchromatic morphologies for NGC 7673. See
Figure 2 for explanation of frames.}

\figcaption[]{Spectral Energy Distribution for NGC 3310 and
NGC 3351.}

\figcaption[]{Spectral Energy Distribution for NGC 3690 and
NGC 3991.}

\figcaption[]{Spectral Energy Distribution for NGC 4861 and
NGC 7673.}

\figcaption[]{The UV spectral slope ($\omega$) vs. the
ratio of the UV spectral slope to the FIR slope $\omega$ / 
$\alpha$.  NGC 3690 stands out as being a galaxy with a very
large FIR slope compared to the other starbursts.}

\clearpage

\setcounter{figure}{0}

\begin{figure}
\plotfiddle{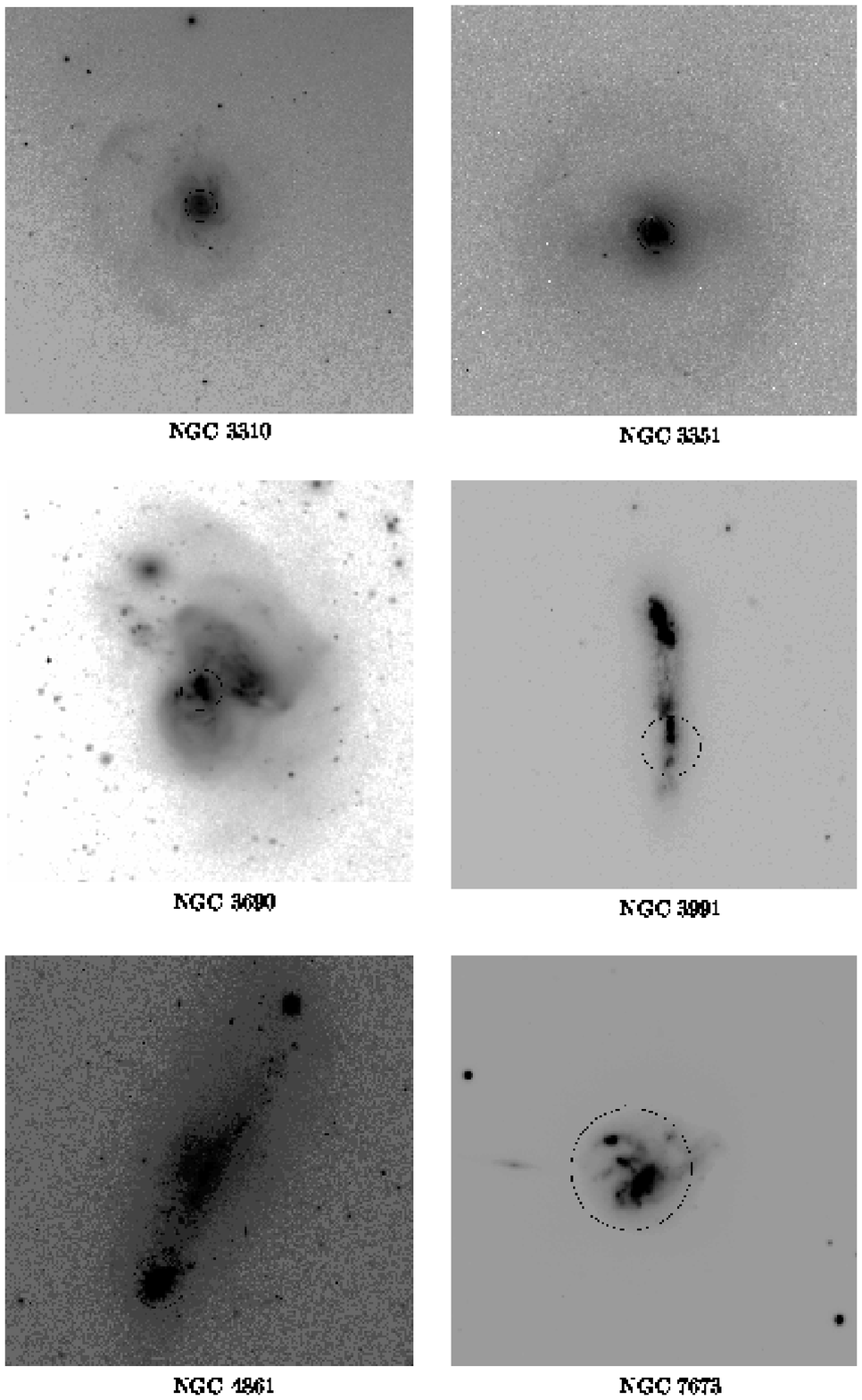}{6.0in}{0}{80}{80}{-250}{-100}
\caption{}
\end{figure}

\clearpage

\begin{figure}
\plotfiddle{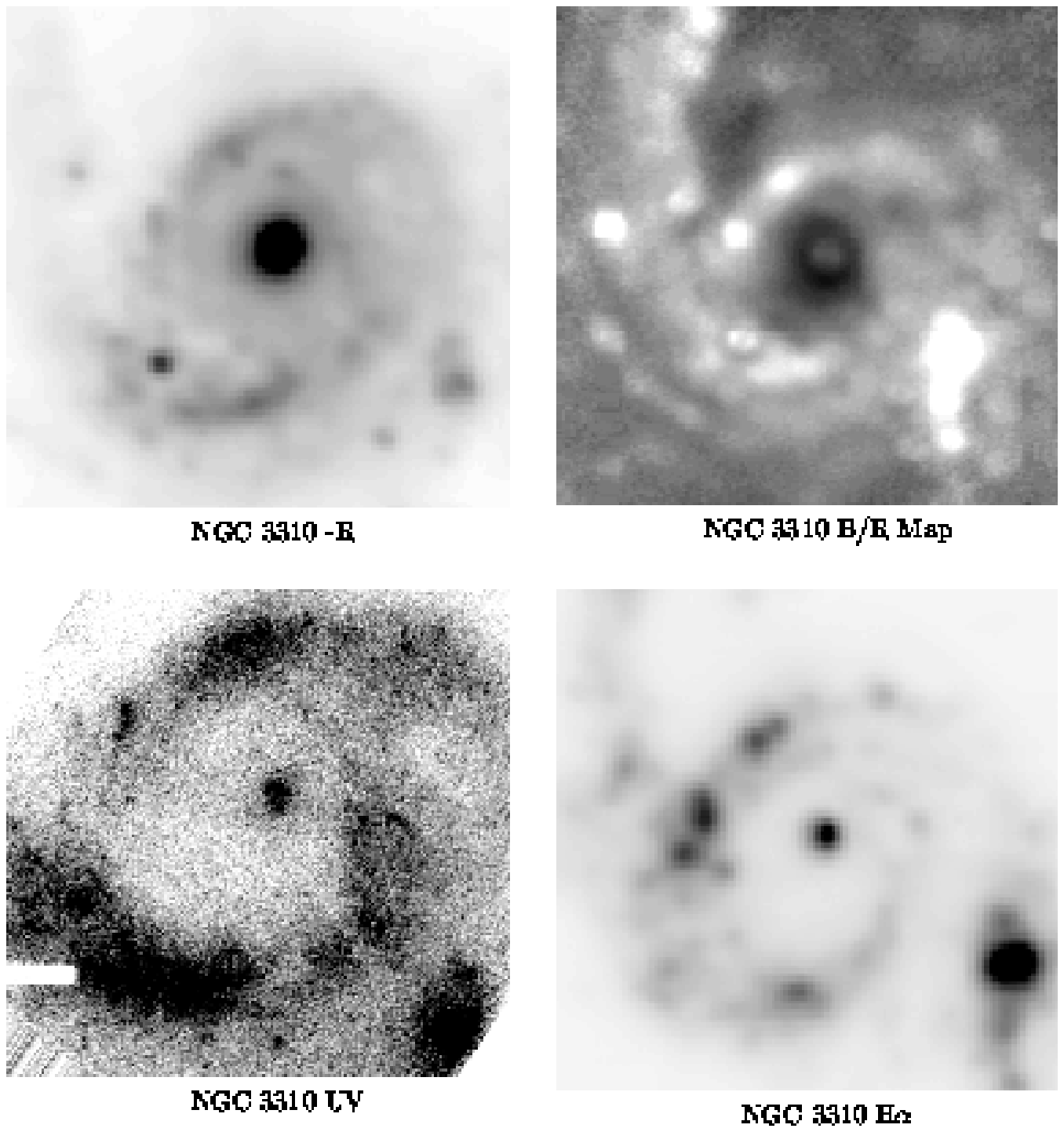}{6.0in}{0}{80}{80}{-250}{-100}
\caption{}
\end{figure}

\clearpage

\begin{figure}

\plotfiddle{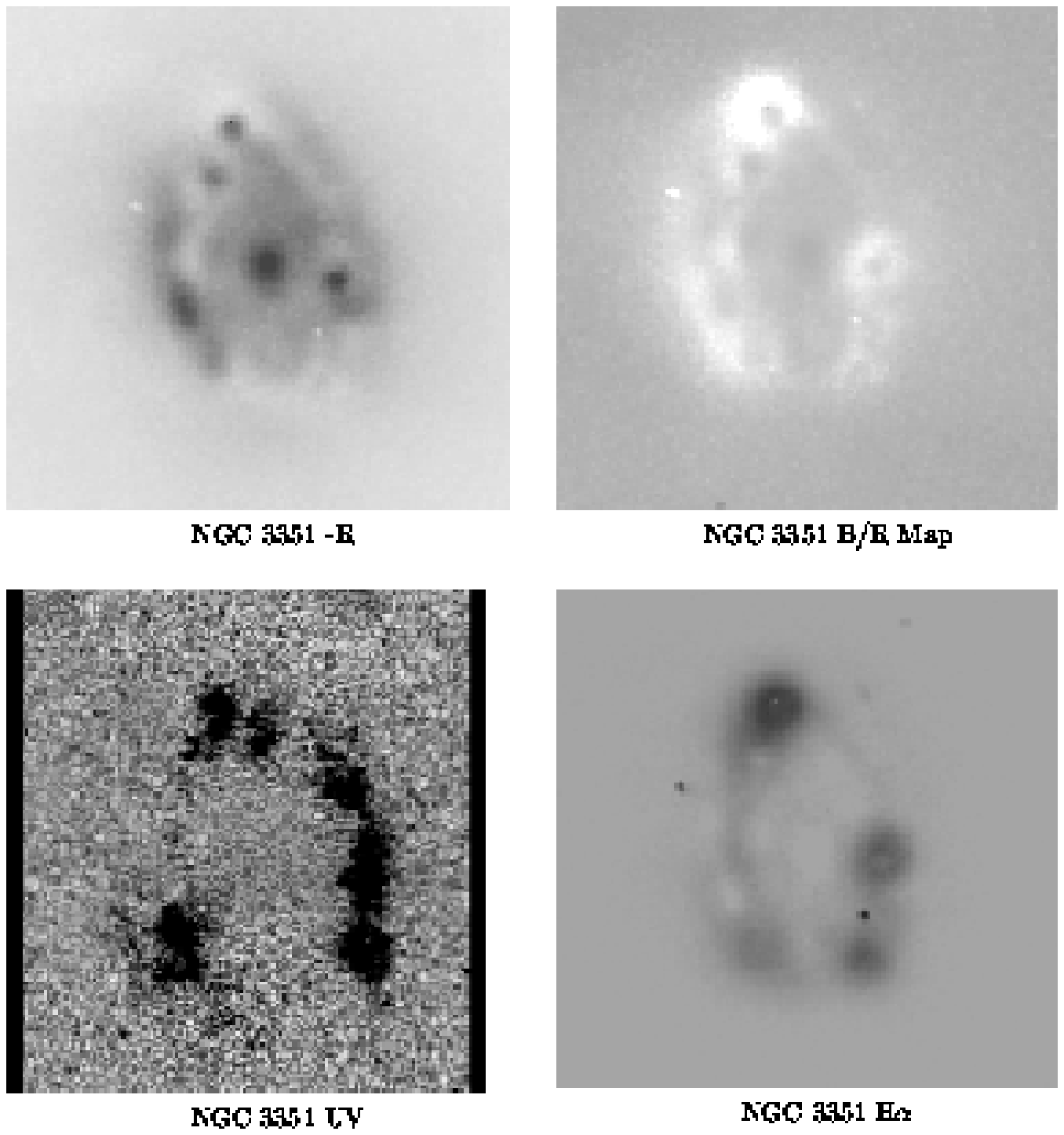}{6.0in}{0}{80}{80}{-250}{-100}
\caption{}
\end{figure}

\clearpage

\begin{figure}
\plotfiddle{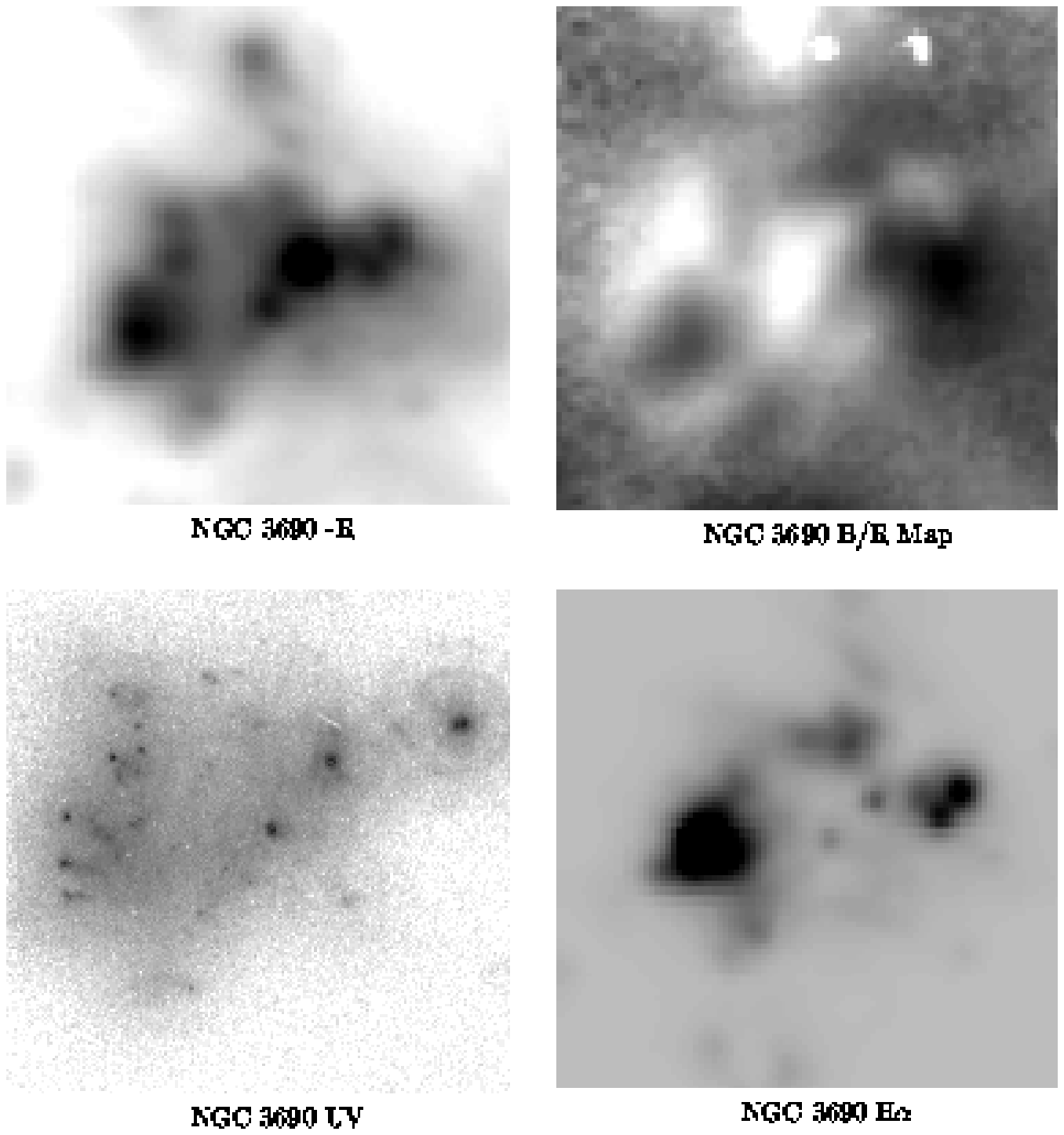}{6.0in}{0}{80}{80}{-250}{-100}
\vskip 1.0in
\caption{}
\end{figure}

\clearpage

\begin{figure}
\plotfiddle{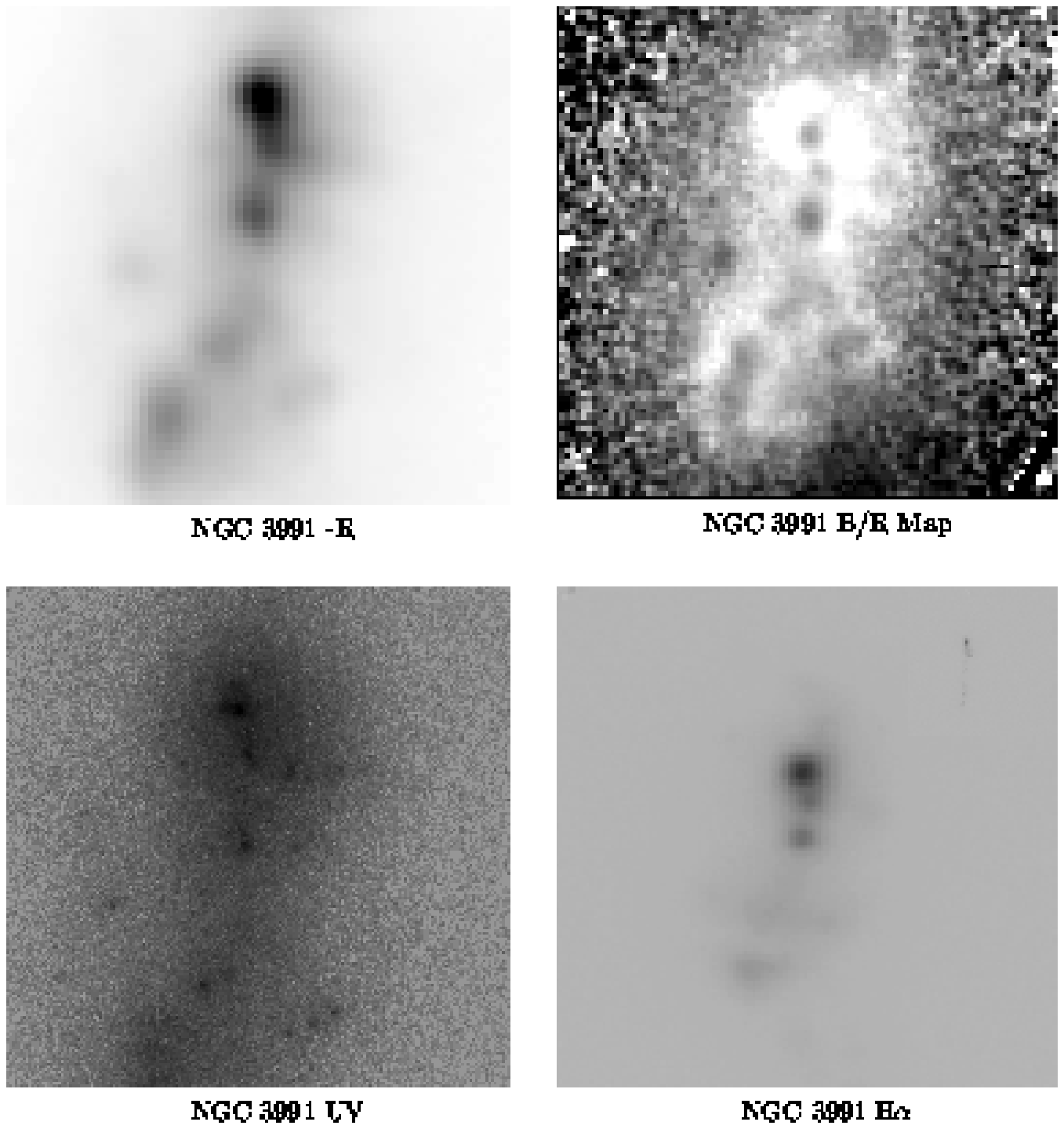}{6.0in}{0}{80}{80}{-250}{-100}
\vskip 1in
\caption{}
\end{figure}

\clearpage

\begin{figure}
\plotfiddle{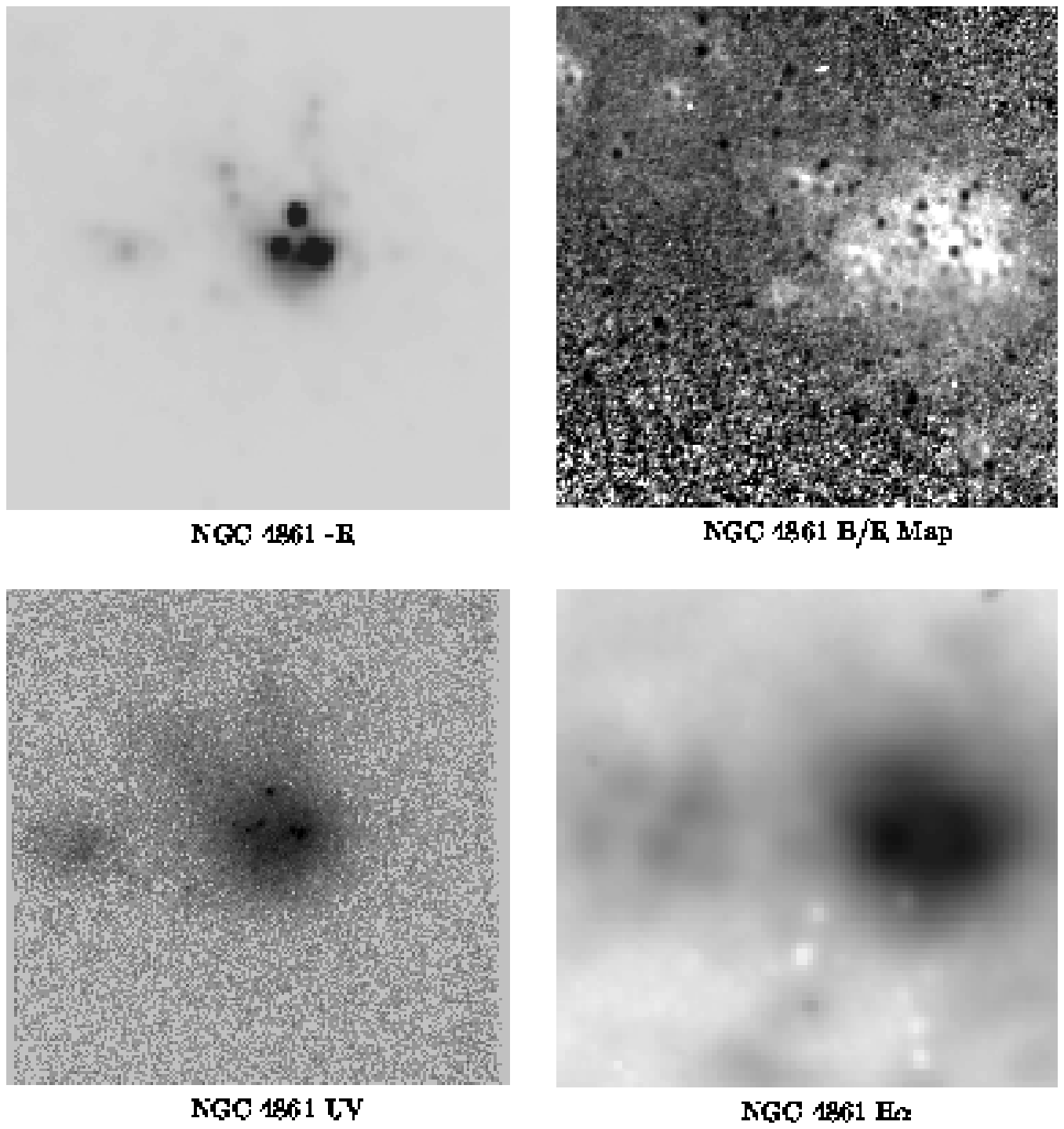}{6.0in}{0}{80}{80}{-250}{-100}
\vskip 1in
\caption{}
\end{figure}

\clearpage

\begin{figure}
\plotfiddle{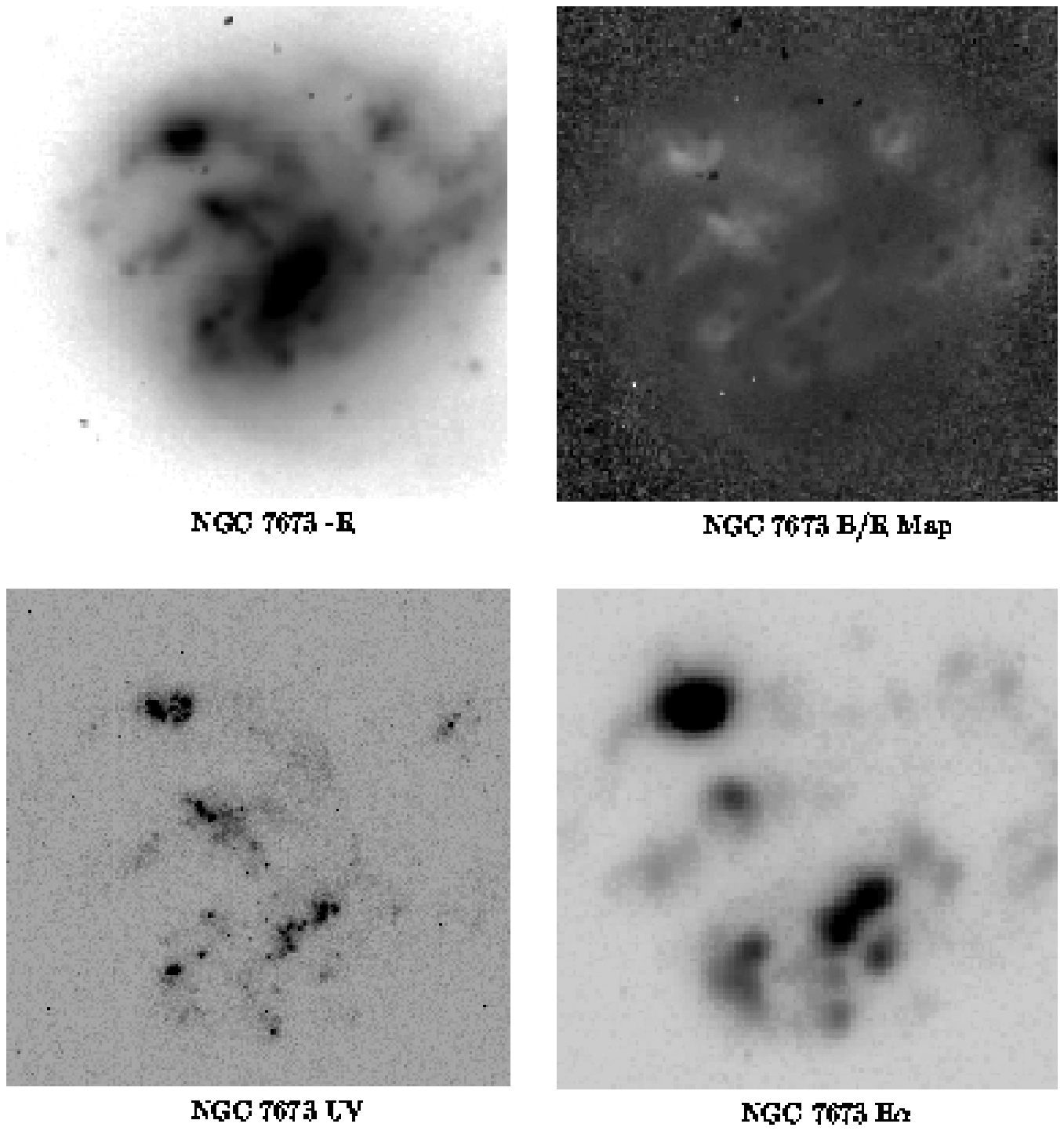}{6.0in}{0}{80}{80}{-250}{-100}
\caption{}
\end{figure}

\clearpage

\begin{figure}
\plotfiddle{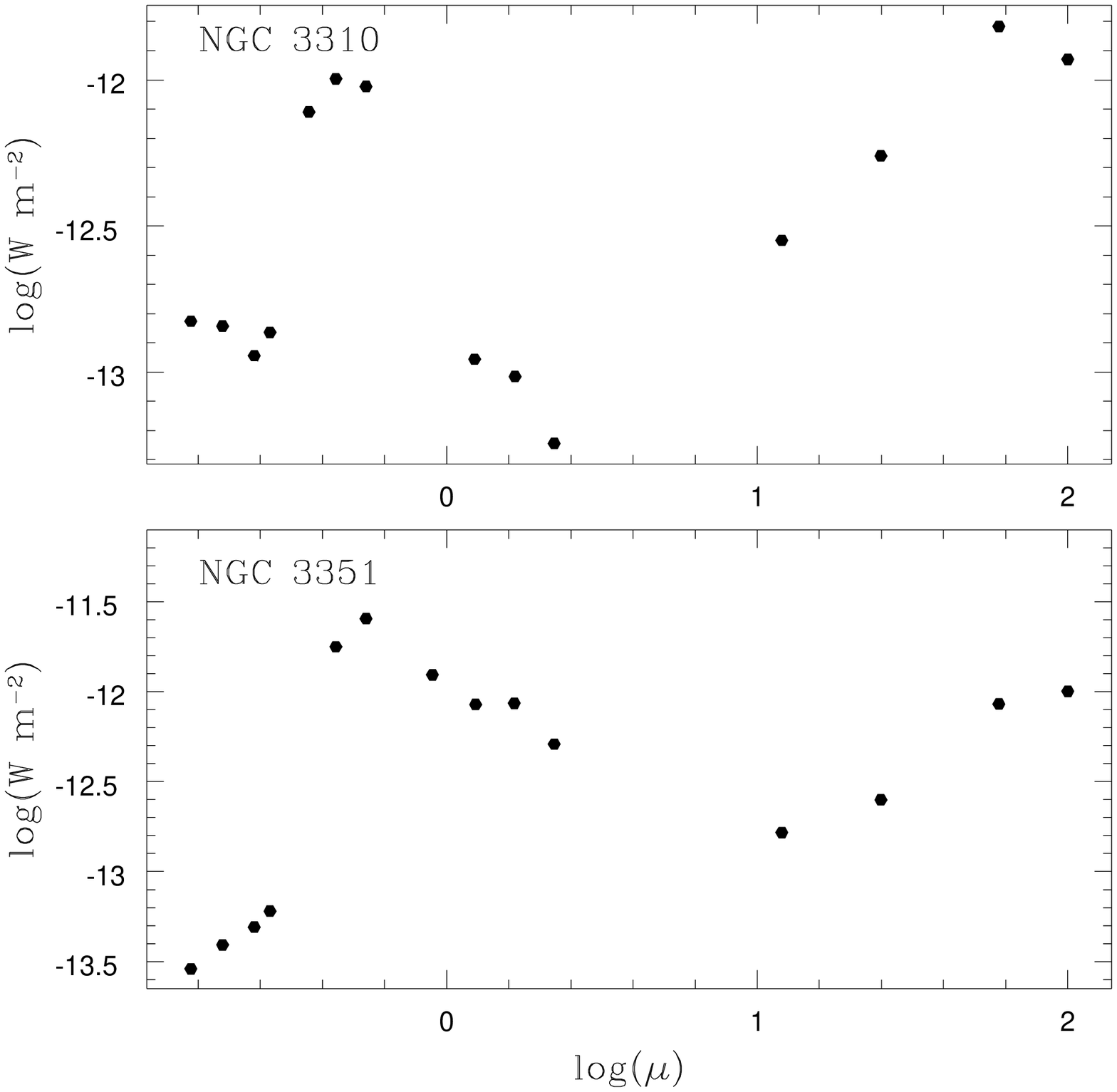}{6.0in}{0}{80}{80}{-250}{-100}

\caption{}
\end{figure}

\clearpage

\begin{figure}
\plotfiddle{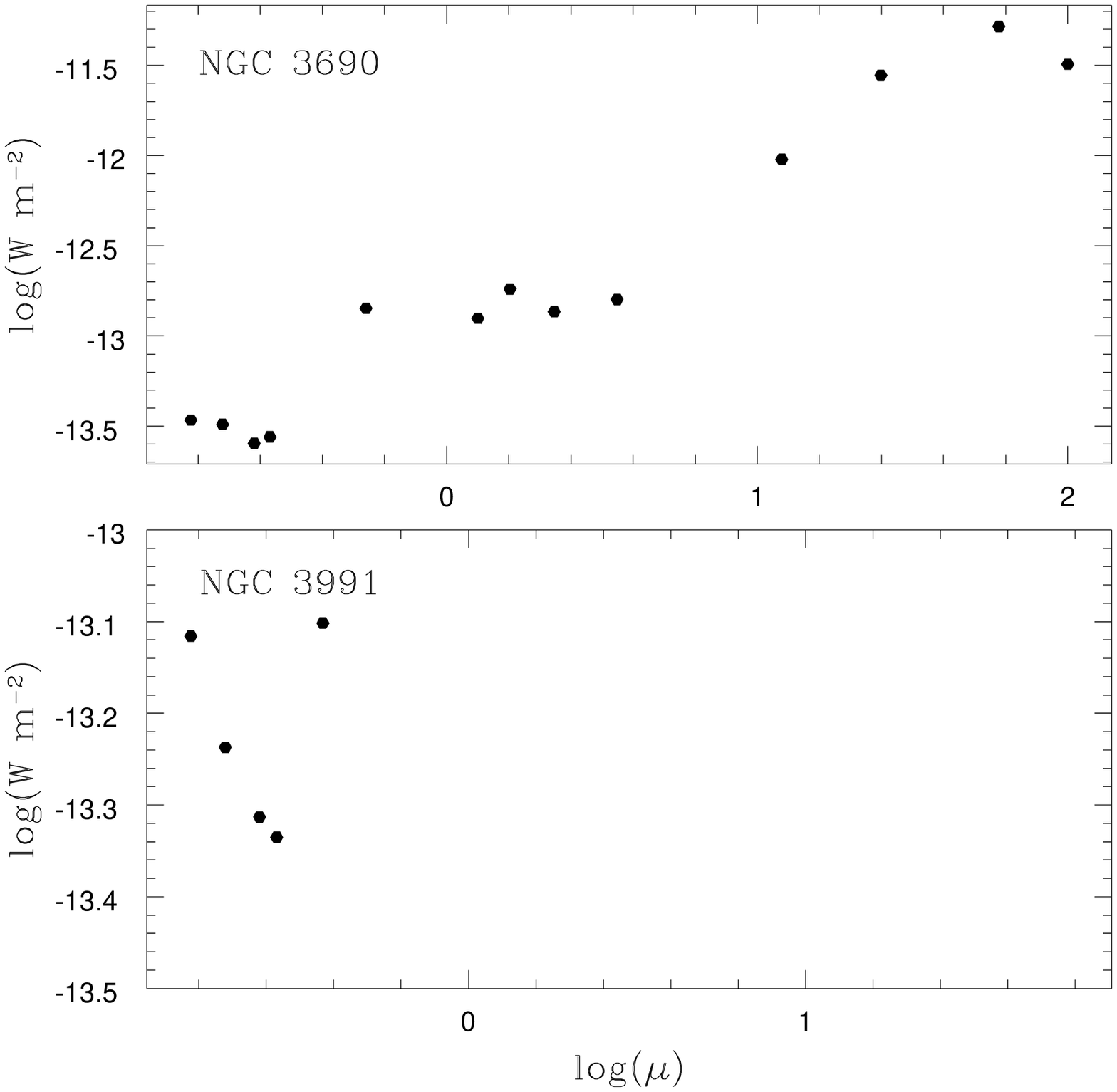}{6.0in}{0}{80}{80}{-250}{-100}
\caption{}
\end{figure}

\clearpage

\begin{figure}
\plotfiddle{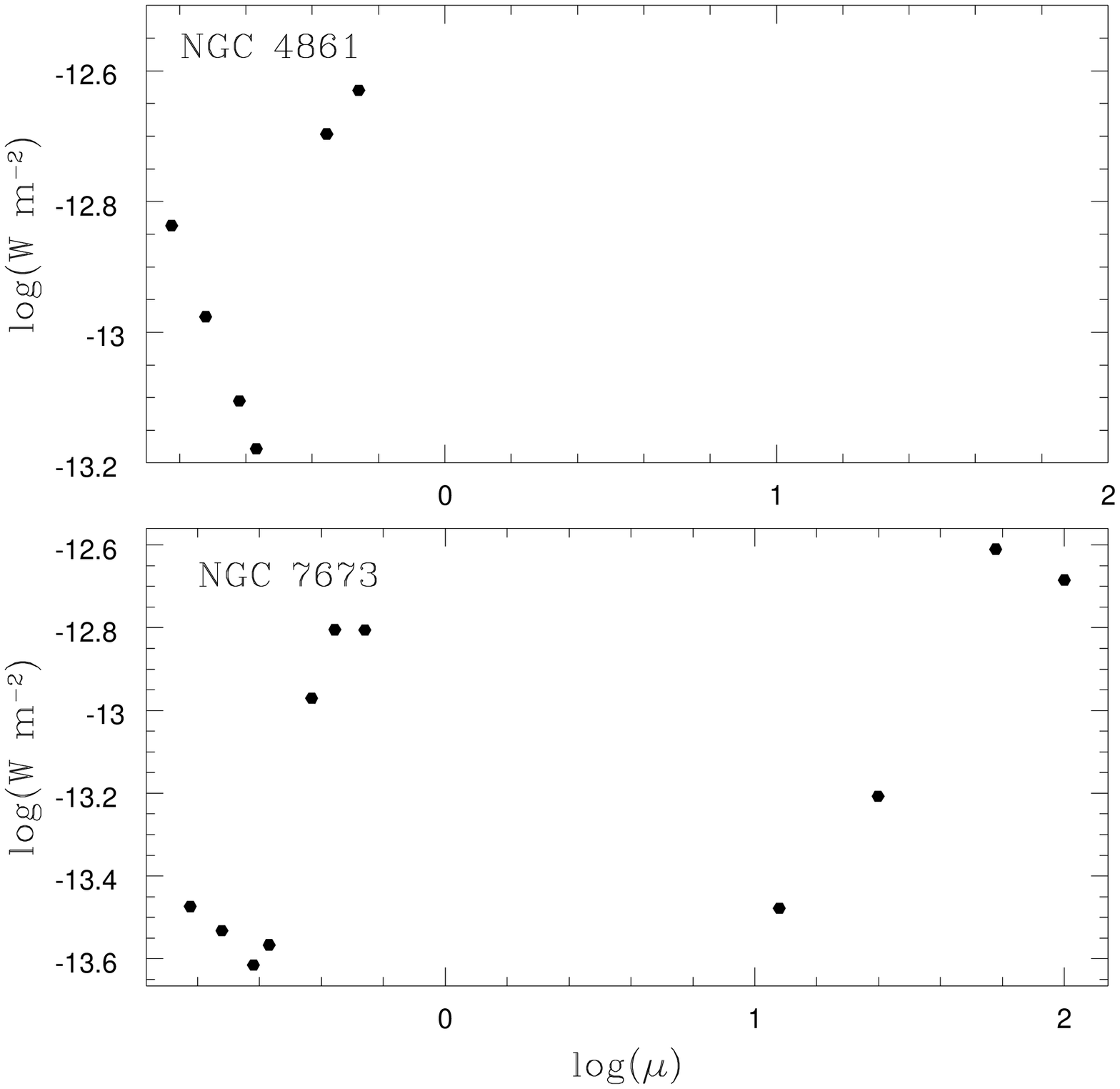}{6.0in}{0}{80}{80}{-250}{-100}
\caption{}
\end{figure}

\clearpage

\begin{figure}
\plotfiddle{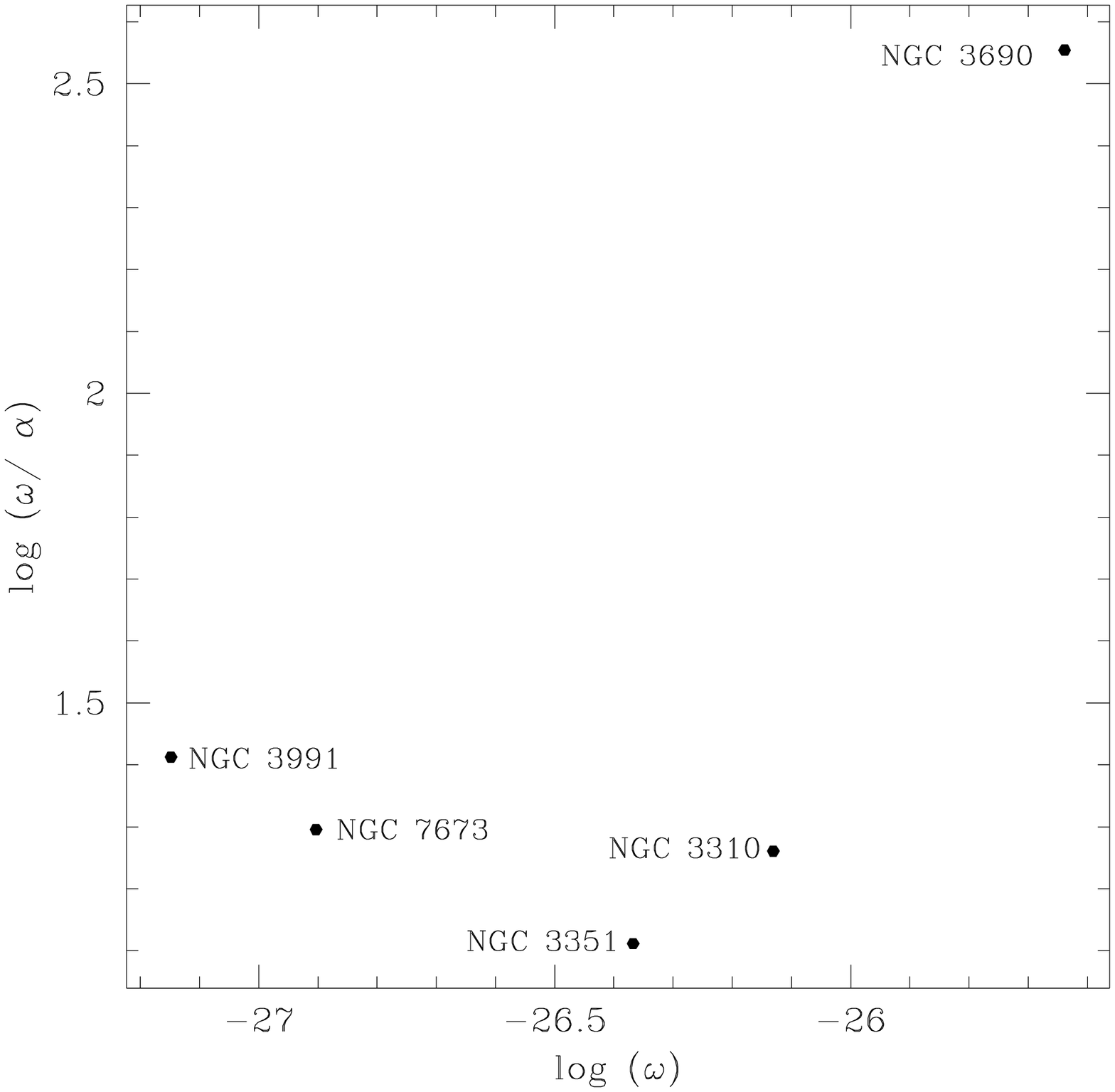}{6.0in}{0}{80}{80}{-250}{-100}
\caption{}
\end{figure}

\end{document}